\documentclass[pra,nofootinbib,showpacs,showkeys,preprintnumbers,amsmath,amssymb]{revtex4-2}

\usepackage{graphicx}
\usepackage{dcolumn}
\usepackage{bm}
\usepackage[usenames]{color}
\usepackage{color}
\usepackage{graphicx,epsfig}
\usepackage{pstricks}

\usepackage[colorlinks={true}]{hyperref}
\hypersetup{colorlinks=true,linkcolor=red,citecolor=blue,urlcolor=blue}

\begin{document}
\title{
Quantum correlations in general qubit-qudit axially symmetric states
}

\author{Saeed~Haddadi\footnote{haddadi@ipm.ir}}
\affiliation{School of Particles and Accelerators, Institute for Research in Fundamental Sciences (IPM),\\ P.O. Box 19395-5531, Tehran, Iran}

\author{Elena~I.~Kuznetsova\footnote{kuznets@icp.ac.ru}}
\affiliation{
Federal Research Center of Problems of Chemical Physics and Medicinal Chemistry,
Russian Academy of Sciences, Chernogolovka 142432, Moscow Region, Russia}

\author{M.~A.~Yurischev\footnote{yur@itp.ac.ru}}
\affiliation{
Federal Research Center of Problems of Chemical Physics and Medicinal Chemistry,
Russian Academy of Sciences, Chernogolovka 142432, Moscow Region, Russia}

\date{\today}

\begin{abstract}
\textbf{Abstract.} The development and improvement of analytical methods for evaluating nonclassical correlations is one of the most important tasks in quantum information science. In this paper, we investigate a mixed spin-$(1/2,S)$ system with an arbitrary spin $S$, where the interactions satisfy the U(1) axial symmetry. Analytical formulas for the local quantum uncertainty (LQU) and local quantum Fisher information (LQFI) are derived directly from the elements and eigenvalues of the density matrix. These results are then used to conduct a comparative analysis of the discord-like quantum correlations, LQU and LQFI, in the system at thermal equilibrium. The high-temperature asymptotics of both quantum correlations are found explicitly. Despite the destructive role of temperature in general, the calculations show that the quantum correlations can increase with temperature in local intervals. Under certain conditions, temperature even generates quantum correlations from uncorrelated ground states. Further, as the system cools, quantum correlations can undergo a series of abrupt transitions with a smooth temperature change. These phenomena are demonstrated for different choices of coupling parameters and spin lengths $S$.
\end{abstract}

\medskip 


\maketitle

\section{Introduction}
\label{sect:Intro}
Quantum information technologies are rapidly entering all aspects of human life.
The current state of quantum computing is called the Noisy Intermediate-Scale Quantum
(NISQ) era \cite{P18,BCK22,AE23}.
Today's quantum computers are limited to hundreds of qubits and have relatively short
coherence times and significant errors \cite{BCK22,CBB23}.
In general, they are far from universal, fault-tolerant quantum computers, which are
believed to make it possible to effectively simulate any processes occurring in Nature.
Nevertheless, existing quantum computers can already solve important real-world
problems.

It is believed that the speedup of quantum computers is due to quantum correlations,
although there is no rigorous evidence.
Until the 21st century, the quantum entanglement was considered to be the only
ingredient of quantum properties leading to such correlations.
It emerges in various phenomena and protocols, including the Einstein-Podolsky-Rosen thought experiment, Bell's inequality tests, superdense coding, teleportation, quantum cryptography, and so
on \cite{Z98,NC00}.

 Many measures of quantum correlations have been proposed to date, with their properties now being extensively investigated through both theoretical and experimental approaches \cite{AFOV08,HHHH09,MBCPV12,AFY14,Z17,HB18,KASSVSH21}. Quantum correlations are characterized as the difference between total and classical correlations \cite{Henderson2001,Ollivier2001}.
Besides entanglement, the most important among them are quantum discord, one-way
quantum work deficit and some others \cite{ABC16,BDSRSS18}.
The idea of quantum discord is based on local measurements and optimization that
were applied to the mutual information.
This concept was subsequently extended to various physical and informational quantities, leading to the introduction of some discord-type measures of quantum correlations. Notable examples include geometric quantum discord, measurement-induced disturbance, and so forth \cite{MBCPV12,ABC16,BDSRSS18}.
The local quantum uncertainty (LQU) and local quantum Fisher information (LQFI), based
respectively on the Wigner-Yanase and quantum Fisher information,
belong to this type of measures \cite{GTA13,GSGTFSSOA14,KLKW18}.

For pure quantum states, quantum discord and entanglement coincide. However, unlike entanglement, discord can persist in separable mixed states, where entanglement is strictly zero. Notably, the set of separable states occupies a finite volume within the system’s Hilbert space \cite{ZHSL98}, whereas the set of zero-discord states is vanishingly small \cite{FACCA10}. This fundamental distinction highlights a key difference between discord and entanglement. Furthermore, Datta et al. \cite{D08,DSC08} evaluated discord in the Knill-Laflamme DQC1 model \cite{KL98}, demonstrating that it scales with quantum efficiency, even as entanglement remains negligible throughout the computation.

The two-site spin-$(1/2,S)$ models naturally arise after tracing out all degrees of
freedom from a many-body system except for two spins.
On the other hand, there are heterodinuclear complexes with pronounced dimer magnetic
structures.
For instance, the crystalline compound [Ni(dpt)(H$_2$O)Cu(pba)]$\cdot$2H$_2$O, where
dpt=bis-(3-aminopropyl)amine and pba=1,3-propylenebis(oxamato), is regarded as the
$s=1/2$ and $S=1$ ferrimagnetic dimer \cite{HNMTK99}.
The copper(II)-chromium(III) complex [Cu(oxpn)Cr(5-Cl-phen)$_2$](NO$_3$)$_3$ with
N,N$^\prime$-bis(3-aminopropyl)oxamide and 5-chloro-1,~10-phenanthroline consists of
(1/2,\,3/2) clusters with  ferromagnitic spin-exchange couplings \cite{PP17}.
A hybrid system $(1/2,S)$ with $S=1$ is realized in deuterium atoms in which hyperfine
splitting is caused by the interaction of nuclear and electron magnetic moments
\cite{APB60,WR72,KHHZK20}.
Besides, spin-orbit coupling of electron spin with its orbital angular momentum leads
to the $(1/2,S)$ system in which ``spin'' $S$ plays a role of the orbital angular
momentum (here a qubit-qudit state is realized on {\em one} particle).
There are also specific compounds containing a stable isotope, such as $^{169}$Tm, with
nuclear spin 1/2 and isotopes $^{17}$O or $^{73}$Ge, whose nuclear spins are 5/2 and
9/2, respectively.
Another example is the hydroxyl radicals $\cdot$OH, which are very stable in the form
of a rarefied gas.

The two-site spin-$(1/2,S)$ models have been considered in a number of works.
In particular, SU(2)-invariant states were studied in Refs.~\cite{S03,S05}.
These states are invariant under uniform rotation of both spins, which means that the
density matrix commutes with all components of the total spin.
For it, many important nonclassical correlations, such as quantum entanglement in the
form of negativity \cite{S03,S05}, quantum entanglement of formation \cite{MC08},
relative entropy of entanglement \cite{WW08}, quantum discord \cite{CG13}, one-way
deficit \cite{WMFW14} and local quantum uncertainty \cite{FA15} have been found.
Moreover, the Holevo quantity has been also obtained in \cite{WGFW22}.
Unfortunately, SU(2)-invariant states are single-parameter and in fact correspond to
only one model--a fully isotropic Heisenberg dimer.

In this paper, we consider $(1/2,S)$-systems with lower symmetry, namely with the axial
symmetry, that is, when the density matrix commutes with only one component of the
total spin.
Unlike the SU(2) case, the axially
symmetric state contains $8S+1$ real independent parameters (see
Sec.~\ref{sect:rho}) and covers many physically important systems such as
Heisenberg XXZ model in an inhomogeneous external magnetic field with
Dzyaloshinskii-Moriya (DM) interaction, various cases of single- and two-ion
anisotropies, etc.
For such systems, the quantum entanglement, quantified by negativity, was investigated
in Refs.~\cite{LRKL12,VS21,VST22}.

Many closely related works in the literature have explored quantum correlations and the influence of interaction terms on quantum systems. For instance,  Erol et al. \cite{Erol2014} analyzed various entanglement measures and the local operations and classical communication (LOCC)-maximized quantum Fisher information in two-qubit systems, providing insights into quantum estimation theory. Ozaydin and  Altintas \cite{Ozaydin2015} demonstrated how DM interaction can enhance quantum metrology by surpassing the shot-noise limit, highlighting its significance for precision measurements. Slaoui et al. \cite{Slaoui2019} conducted a comparative study between quantum correlations quantifiers LQFI and LQU in the Heisenberg isotropic (and anisotropic) XY model subjected to an external magnetic field, showing that LQU is majorized by LQFI in any metrological task of the phase estimation. Haseli \cite{Haseli2020} investigated LQFI and LQU in a two-qubit Heisenberg XYZ chain with DM interaction, shedding light on their impact on quantum uncertainty quantifiers. Lian and Liu \cite{Lian2021} extended the study of quantum Fisher information to a qubit-qutrit system in the context of Garfinkle–Horowitz–Strominger dilaton space-time, exploring relativistic effects on quantum estimation. Fedorova and Yurischev \cite{Fedorova2022} examined quantum discord, LQU and LQFI in a two-spin-1/2 Heisenberg chain with both DM and Kaplan–Shekhtman–Entin-Wohlman–Aharony (KSEA) interactions, emphasizing the temperature effects and system parameters. Additionally, Mahdavifar et al. \cite{Mahdavifar2024} studied the resilience of quantum spin fluctuations under DM interaction, revealing its role in stabilizing quantum correlations in spin systems. Also, a study of quantum correlations in hybrid $(1/2,1)$-systems was carried out using negativity, as well as LQU and LQFI \cite{BARDAH23}.

Compared to our recent papers \cite{Yurischev20231,Yurischev2023,Yurischev2025}, which focused on deriving compact closed-form expressions for LQU and LQFI in general qubit-qubit X states and qubit-qutrit axially
symmetric states, our present work extends the analysis to a more generalized mixed spin-$(1/2,S)$ system with arbitrary spin $S$. While the previous study \cite{Yurischev2025} specifically examined a spin-$(1/2,1)$ system with ten independent interaction parameters, the current work formulates analytical expressions for LQU and LQFI in a broader class of systems satisfying U(1) axial symmetry, offering a more comprehensive framework for evaluating nonclassical correlations in a hybrid spin-$(1/2,S)$ system.

The significance and novelty of our present paper lie in its generalized formulation, which allows for the study of arbitrary spin-$(1/2,S)$ systems, thereby broadening the applicability of LQU and LQFI as tools for characterizing quantum correlations. The analytical expressions derived for these measures enable a direct and efficient evaluation of quantum correlations from the density matrix entries and eigenvalues, simplifying their computation in a wide range of physical systems. The discovery of temperature-induced quantum correlations and multiple abrupt transitions highlights new aspects of thermal quantum effects that could be relevant for quantum thermodynamics and quantum information processing. By systematically analyzing how quantum correlations behave across different temperature regimes and spin configurations, this study provides deeper physical insights into the interplay between thermal fluctuations and quantum coherence, offering potential applications in quantum materials and quantum technologies.

The present paper is structured as follows.
In the next section, the general form of the axially
symmetric density matrix is
established.
In Sec.~\ref{sec:Corr}, we derive formulas that are suitable for actually calculating LQU and
LQFI correlations.
The spin-(1/2,$S$) system at thermal equilibrium is considered in Sec.~\ref{sec:appl}.
Sec.~\ref{sec:res} includes high-temperature expansions of quantum correlations,
numerical results and discussion.
Finally, concluding remarks in Sec.~\ref{sec:Summ} close the paper.


\section{
A form of axially symmetric states
}
\label{sect:rho}
Take a system composed of two particles with spins 1/2 and $S$.
Let its quantum state be invariant under transformations of the axial
symmetry group
U(1), consisting of rotations $R_z(\phi) = \exp(-i\phi{\cal S}_z)$ around the $z$-axis
on angle $\phi\in[0,\pi)$.
This means that the density matrix commutes with the $z$-component of total
spin, given by
\begin{equation}
   \label{eq:Sz_total}
   {\cal S}_z=s_z\otimes I_{2S+1}+I_2\otimes S_z
	 ={\rm diag}\,[1/2+S,1/2+S-1,\ldots,1/2-S,-1/2+S,\ldots,-1/2-S+1,-1/2-S],
\end{equation}
where $I_2$ and $I_{2S+1}$ are the identity operators of the second and ($2S+1$)-th
orders, respectively.
The matrix (\ref{eq:Sz_total}) of $2(2S+1)$-th order is diagonal, and let its
eigenvalues be numbered from 0 to $4S+1$.
Except for the extreme eigenvalues $\pm(1/2+S)$, the remaining $4S$ internal
eigenvalues of the matrix (\ref{eq:Sz_total}) are two-fold degenerate: the $k$-th and
$(2S+k)$-th ones are equal to each other ($k=1,\ldots,2S$).

The most general Hermitian matrix which commutes with the $z$-component of total spin
(\ref{eq:Sz_total}) can be written as
\begin{equation}
   \label{eq:rho}
   \rho=
	 \left(
      \begin{array}{clccccccrc}
      p_0&\ &\ &\ &\ &\ &\ &\ &\ &\ \\
      \ &a_1&0&\ldots&0&u_1&0&\ldots&0&\ \\
      \ &0&a_2&\ldots&0&0&u_2&\ldots&0&\ \\
      \ &\vdots&\vdots&\ddots&\vdots&\vdots&\vdots&\ddots&\vdots&\ \\
      \ &0&0&\ldots&a_{2S}&0&0&\ldots&u_{2S}&\ \\
      \ &u_1^*&0&\ldots&0&a_{2S+1}&0&\ldots&0&\ \\
      \ &0&u_2^*&\ldots&0&0&a_{2S+2}&\ldots&0&\ \\
      \ &\vdots&\vdots&\ddots&\vdots&\vdots&\vdots&\ddots&\vdots&\ \\
      \ &0&0&\ldots&u_{2S}^*&0&0&\ldots&a_{4S}&\ \\
			\ &\ &\ &\ &\ &\ &\ &\ &\ &p_{4S+1}
      \end{array}
   \right).\
\end{equation}
This matrix has a block-diagonal structure
$(1\times1)\oplus(4S\times4S)\oplus(1\times1)$, where the interior $4S\times4S$
subblock is sparse, and its nonzero entries are located only on the main diagonal and
two sub-diagonals.
Positive definiteness is assumed for any density matrices ($\rho\ge0$), as well as
normalization of the trace to unity (${\rm Tr}\rho=1$).
In the case under consideration, this means that $p_0,a_1,\ldots,a_{4S},p_{4S+1}\ge0$,
$a_ka_{2S+k}\ge|u_k|^2$ and $p_0+p_{4S+1}+\sum_{i=1}^{4S}a_i=1$.
Below, for brevity, the density matrices like (\ref{eq:rho}) will be called as the axially
symmetric
states.
Obviously, the density matrix (\ref{eq:rho}) contains $8S+1$ free real parameters.

The transformation $R$ diagonalizing $\rho$ is presented in Appendix~\ref{appen:A}.
It allows us to extract the eigenvalues of the interior subblock $4S\times4S$ of the
density matrix (\ref{eq:rho}), which are given as
\begin{equation}
   \label{eq:p_k}
   p_{k,2S+k}=\frac{1}{2}\Big(a_k+a_{2S+k}\pm\sqrt{(a_k-a_{2S+k})^2+4|u_k|^2}\Big).
\end{equation}

Now, we proceed to the derivation of formulas for the quantum correlations LQU and LQFI.


\section{
Quantum correlations for axially symmetric states
}
\label{sec:Corr}

\subsection{LQU}
\label{subsec:LQU}
The total uncertainty of an observable $H$ in the quantum-mechanical state $\rho$ is
usually expressed by the variance
\begin{equation}
   \label{eq:var}
   {\rm Var}(\rho,H)=\langle H^2\rangle_{\rho}-\langle H\rangle_{\rho}^2.
\end{equation}
On the other hand, the quantum contribution to the total statistical error may be
reliable quantify via the Wigner-Yanase skew information \cite{WY63,L03,L03a}
\begin{equation}
   \label{eq:WY}
   {\cal I}(\rho,H)=-\frac{1}{2}{\rm Tr}[\sqrt{\rho},H]^2,
\end{equation}
where $[.,.]$ denotes the commutator.
Notice that the skew information (\ref{eq:WY}) is not grater than the variance
(\ref{eq:var}), namely
\begin{equation}
   \label{eq:WY-V}
   {\cal I}(\rho,H)\le{\rm Var}(\rho,H),
\end{equation}
where the equality is achieved for pure states when classical ignorance does not occur
\cite{GTA13}.
This makes it possible to introduce a discord-type measure $\cal U$ (also called LQU)
of quantum correlations in any bipartite system $AB$ as follows \cite{GTA13}
\begin{equation}
   \label{eq:Udef}
   {\cal U}(\rho)=\min_{H_A}{\cal I}(\rho,H_A),
\end{equation}
in which the minimum is taken over all local observables $H_A$ on the subsystem $A$.
It is worth mentioning that the LQU is a genuine quantifier of quantum correlations,
and it has been shown that the LQU meets all the physical conditions of a measure of
quantum correlations.

Importantly, the authors \cite{GTA13} were able to perform optimization for qubit-qudit
systems and presented the measure (\ref{eq:Udef}) in the following form
\begin{equation}
   \label{eq:Um}
   {\cal U}=1-\lambda_{\max}^{(W)},
\end{equation}
where $\lambda_{\max}^{(W)}$ denotes the maximum eigenvalue of the $3\times3$
symmetric matrix $W$ whose entries are
\begin{equation}
   \label{eq:W}
   W_{\mu \nu}={\rm Tr}\{\rho^{1/2}(\sigma_\mu\otimes{\rm I}_{2S+1})\rho^{1/2}(\sigma_\nu\otimes{\rm I}_{2S+1})\}
\end{equation}
with $\mu,\nu=x,y,z$ and $\sigma_{x,y,z}$ are the Pauli matrices.
In principle, this opens up the possibility of analytical calculating LQU by solving
secular (algebraic) equation of the third degree for the matrix $W$.

However, passing under the trace symbol in Eq.~(\ref{eq:W}) to the diagonal
representation of the density matrix (\ref{eq:rho}) and using
Eqs.~(\ref{eq:sxI})--(\ref{eq:szI}) from Appendix~\ref{appen:B}, we find that the
matrix $W$ for axially
symmetric states is diagonal, $W_{yy}=W_{xx}$ and
\begin{eqnarray}
   \label{eq:Wxx}
   &&\frac{1}{2}W_{xx}={\tilde q}_1^2\sqrt{p_0p_{2S+1}}+{\tilde q}_{2S}^2\sqrt{p_{2S}p_{4S+1}}
	 +|{\tilde u}_1|^2\sqrt{p_0p_1}+|{\tilde u}_{2S}|^2\sqrt{p_{4S}p_{4S+1}}
   \nonumber\\
   &&+\sum_{k=1}^{2S-1}\big[{\tilde q}_k^2{\tilde q}_{k+1}^2\sqrt{p_kp_{2S+1+k}}
	 +{\tilde q}_k^2|{\tilde u}_{k+1}|^2\sqrt{p_kp_{k+1}}
   +|{\tilde u}_k|^2{\tilde q}_{k+1}^2\sqrt{p_{2S+k}p_{2S+1+k}}
   +|{\tilde u}_k|^2|{\tilde u}_{k+1}|^2\sqrt{p_{k+1}p_{2S+k}}\big]
\end{eqnarray}
and
\begin{equation}
   \label{eq:Wzz}
   W_{zz}=p_0+p_{4S+1}+\sum_{k=1}^{2S}\big[(p_k+p_{2S+k})({\tilde q}_k^2-|{\tilde u}_k|^2)^2
	 +8{\tilde q}_k^2|{\tilde u}_k|^2\sqrt{p_kp_{2S+k}}\big].
\end{equation}
Therefore, LQU would be given by
\begin{equation}
   \label{eq:U0U1}
   {\cal U}=\min{\{{\cal U}_0,{\cal U}_1\}},
\end{equation}
where two branches (sub-functions) are ${\cal U}_0=1-W_{zz}$ and ${\cal U}_1=1-W_{xx}$.

Substituting expressions (\ref{eq:q_ku2_k}) into Eqs.~(\ref{eq:Wxx}) and
(\ref{eq:Wzz}), we arrive at the final formulas for the branches
\begin{equation}
   \label{eq:U0}
   {\cal U}_0=\sum_{k=1}^{2S}\Bigg[(\sqrt{p_k}-\sqrt{p_{2S+k}})^2-\frac{(a_k-a_{2S+k})^2}{(\sqrt{p_k}+\sqrt{p_{2S+k}})^2}\Bigg]
\end{equation}
and
\begin{equation}
   \label{eq:U1}
   {\cal U}_1=1-2\Bigg[\frac{a_{2S+1}+\sqrt{p_1p_{2S+1}}}{\sqrt{p_1}+\sqrt{p_{2S+1}}}\sqrt{p_0}
	 +\frac{a_{2S}+\sqrt{p_{2S}p_{4S}}}{\sqrt{p_{2S}}+\sqrt{p_{4S}}}\sqrt{p_{4S+1}}
   +\sum_{k=1}^{2S-1}\frac{(a_k+\sqrt{p_k p_{2S+k}})(a_{2S+1+k}+\sqrt{p_{k+1}p_{2S+1+k}})}
	 {(\sqrt{p_k}+\sqrt{p_{2S+k}})(\sqrt{p_{k+1}}+\sqrt{p_{2S+1+k}})}
	 \Bigg].
\end{equation}

\subsection{LQFI}
\label{subsec:LQFI}
Quantum Fisher information (QFI) is the cornerstone of quantum estimation theory
\cite{H76,H82,BC94,LYLW20}.
Suppose there is a unitary evolution of the quantum state
$\varrho=e^{iH\epsilon}\rho e^{-iH\epsilon}$ with some observable $H$, and
the parameter $\epsilon$ needs to be estimated.
Thus, the QFI can be defined as follows
\begin{equation}
   \label{eq:QFI}
   F(\varrho, H)=\frac{1}{4}{\rm Tr}(\varrho L_\epsilon^2),
\end{equation}
where $L_\epsilon$ is the symmetric logarithmic derivative (SLD) operator satisfying
the Lyapunov equation (known from control theory)
\begin{equation}
   \label{eq:SLD}
   \frac{\partial\varrho}{\partial\epsilon}=\frac{1}{2}(\varrho L_\epsilon+L_\epsilon\varrho).
\end{equation}
Then the accuracy of the estimation of $\epsilon$ is limited by the quantum
Rao-Cram$\rm{\acute e}$r inequality \cite{H82}
\begin{equation}
   \label{eq:RC}
   \Delta\epsilon\ge\frac{1}{\sqrt{NF(\varrho,H)}},
\end{equation}
where $N$ is the number of measurements.

Another application of QFI is to use it as a quantifier of quantum correlations.
Girolami et al.~\cite{GSGTFSSOA14} showed that the quantity
\begin{equation}
   \label{eq:Fdef}
   {\cal F}(\rho)=\min_{H_A}F(\rho,H_A),
\end{equation}
where $H_A$ (as in the case of LQU) is a local observable, again satisfies all
the necessary criteria to be defined as a measure of quantum correlations.
Note that here the (local) QFI is minimized, unlike the case of the lower bound
(\ref{eq:RC}).
The measure $\cal F$ is called interferometric power \cite{GSGTFSSOA14} or LQFI
\cite{B14,KLKW18}.
Moreover, if the subsystem $A$ is a qubit, then minimization in (\ref{eq:Fdef}) can be
done and LQFI is given as \cite{GSGTFSSOA14}
\begin{equation}
   \label{eq:calF}
   {\cal F}=1-\lambda_{\max}^{(M)},
\end{equation}
where $\lambda_{\max}^{(M)}$ is the largest eigenvalue of the real symmetric $3\times3$
matrix $M$ with entries \cite{DBA15}
\begin{equation}
   \label{eq:M}
   M_{\mu\nu}=\sum_{{m,n};\ {p_m+p_n\ne0}}\frac{2p_mp_n}{p_m+p_n}\langle m|\sigma_\mu\otimes{\rm I}_{2S+1}|n\rangle
	 \langle n|\sigma_\nu\otimes{\rm I}_{2S+1}|m\rangle.
\end{equation}
From a programming point of view, the right side of this equation represents two nested
loops.

Again using Eqs.~(\ref{eq:sxI})--(\ref{eq:szI}), we find that the matrix $M$ for axially
symmetric
states is also diagonal, $M_{yy}=M_{xx}$, and
\begin{eqnarray}
   \label{eq:Mxx}
   \frac{1}{4}M_{xx}&=&\frac{p_0p_{2S+1}}{p_0+p_{2S+1}}{\tilde q}_1^2
	 +\frac{p_{2S}p_{4S+1}}{p_{2S}+p_{4S+1}}{\tilde q}_{2S}^2
	 +\frac{p_0p_1}{p_0+p_1}|{\tilde u}_1|^2
	 +\frac{p_{4S}p_{4S+1}}{p_{4S}+p_{4S+1}}|{\tilde u}_{2S}|^2
   +\sum_{k=1}^{2S-1}\Big[\frac{p_kp_{2S+1+k}}{p_k+p_{2S+1+k}}{\tilde q}_k^2{\tilde q}_{k+1}^2
   \nonumber\\
	 &+&\frac{p_kp_{k+1}}{p_k+p_{k+1}}{\tilde q}_k^2|{\tilde u}_{k+1}|^2
   +\frac{p_{2S+k}p_{2S+1+k}}{p_{2S+k}+p_{2S+1+k}}|{\tilde u}_k|^2{\tilde q}_{k+1}^2
   +\frac{p_{k+1}p_{2S+k}}{p_{k+1}+p_{2S+k}}|{\tilde u}_k|^2|{\tilde u}_{k+1}|^2\Big]
\end{eqnarray}
and
\begin{equation}
   \label{eq:Mzz}
   M_{zz}=p_0+p_{4S+1}+\sum_{k=1}^{2S}\big[(p_k+p_{2S+k})({\tilde q}_k^2-|{\tilde u}_k|^2)^2
	 +16\frac{p_kp_{2S+k}}{p_k+p_{2S+k}}{\tilde q}_k^2|{\tilde u}_k|^2\big].
\end{equation}
It is interesting to draw our attention to the analogy in the structures, on the one
hand, between Eqs.~(\ref{eq:Wxx}) and (\ref{eq:Mxx}) and, on the other hand, between
Eqs.~(\ref{eq:Wzz}) and (\ref{eq:Mzz}).

As with LQU, this makes it possible to actually calculate LQFI:
\begin{equation}
   \label{eq:F0F1}
   {\cal F}=\min{\{{\cal F}_0,{\cal F}_1\}},
\end{equation}
where the branches (sub-functions) are equal to
${\cal F}_0=1-M_{zz}$ and ${\cal F}_1=1-M_{xx}$.

Substituting (\ref{eq:q_ku2_k}) into Eqs.~(\ref{eq:Mxx}) and
(\ref{eq:Mzz}), we come to a nice formula for the ${\cal F}_0$ branch
\begin{equation}
   \label{eq:F0}
   {\cal F}_0=4\sum_{k=1}^{2S}\frac{|u_k|^2}{a_k+a_{2S+k}}.
\end{equation}
Similarly, the  final formula for the second branch ${\cal F}_1$ can be written as
\begin{eqnarray}
   \label{eq:F1}
   &&{\cal F}_1=1-4\Bigg\{\frac{p_0(a_{2S+1}p_0+p_1p_{2S+1})}{(p_0+p_1)(p_0+p_{2S+1})}
	 +\frac{p_{4S+1}(a_{2S}p_{4S+1}+p_{2S}p_{4S})}{(p_{2S}+p_{4S+1})(p_{4S}+p_{4S+1})}
	 +\frac{1}{2}\sum_{k=1}^{2S-1}\Bigg[
	 \frac{(a_kp_{k+1}+p_kp_{2S+k})p_{k+1}}{(p_k+p_{k+1})(p_{k+1}+p_{2S+k})}
   \nonumber\\
   &&+\frac{a_k\{p_{2S+1+k}[p_{2S+k}(p_k+p_{k+1})+p_kp_{k+1}]+p_kp_{k+1}p_{2S+k}\}+p_kp_{2S+k}(p_kp_{2S+k}-p_{k+1}p_{2S+1+k})}
	 {(p_k+p_{k+1})(p_{k+1}+p_{2S+k})(p_{2S+k}+p_{2S+1+k})(p_{2S+1+k}+p_k)}(a_{2S+1+k}-a_{k+1})
   \nonumber\\
	 &&+\frac{(a_kp_{2S+1+k}+p_kp_{2S+k})p_{2S+1+k}}{(p_k+p_{2S+1+k})(p_{2S+k}+p_{2S+1+k})}
	 \Bigg]
   \Bigg\}.
\end{eqnarray}

Remarkably, both measures ${\cal U}$ and ${\cal F}$ for axially
symmetric states involve only square
radicals and are expressed directly in terms of the matrix elements of the density
matrix (\ref{eq:rho}) and its eigenvalues (\ref{eq:p_k}).
Draw attention that the $2S$ off-diagonal entries $u_k$ appear everywhere under the
modulus symbol.

In contrast to Eqs.~(\ref{eq:Um}), (\ref{eq:W}), (\ref{eq:calF})  and (\ref{eq:M}),
algebraic expressions (\ref{eq:U0}), (\ref{eq:U1}), (\ref{eq:F0}) and (\ref{eq:F1}) for
the branches of quantum correlations are surprisingly simple and very convenient for
programming on a computer.


\section{System at thermal equilibrium}
\label{sec:appl}
As an application of the derived formulas, we consider spin-(1/2,$S$) system in the
state of thermal equilibrium.
Take the Hamiltonian $\cal H$ commuting with ${\cal S}_z$.
It will have an axially symmetrical structure (portrait) like (\ref{eq:rho}) and can be written as
\begin{equation}
   \label{eq:H}
   {\cal H}=
	 \left(
      \begin{array}{clccccccrc}
      E_0&\ &\ &\ &\ &\ &\ &\ &\ &\ \\
      \ &h_1&0&\ldots&0&g_1&0&\ldots&0&\ \\
      \ &0&h_2&\ldots&0&0&g_2&\ldots&0&\ \\
      \ &\vdots&\vdots&\ddots&\vdots&\vdots&\vdots&\ddots&\vdots&\ \\
      \ &0&0&\ldots&h_{2S}&0&0&\ldots&g_{2S}&\ \\
      \ &g_1^*&0&\ldots&0&h_{2S+1}&0&\ldots&0&\ \\
      \ &0&g_2^*&\ldots&0&0&h_{2S+2}&\ldots&0&\ \\
      \ &\vdots&\vdots&\ddots&\vdots&\vdots&\vdots&\ddots&\vdots&\ \\
      \ &0&0&\ldots&g_{2S}^*&0&0&\ldots&h_{4S}&\ \\
			\ &\ &\ &\ &\ &\ &\ &\ &\ &E_{4S+1}
      \end{array}
   \right).\
\end{equation}
This matrix contains $2(4S+1)$ real free parameters (unlike density matrices, since
there is no normalization condition here).
Importantly, the same form is preserved for the powers of $\cal H$, their
sums and hence functions (representable by Tailor series).

Expanding the Hermitian matrix (\ref{eq:H}) into generalized Gell-Mann matrices and
then using the connections of these matrices with the spin ones, we obtain an
expression for the Hamiltonian with $2(4S+1)$ terms of the spin matrices.
Instead of this formal approach, one can also construct the same Hamiltonian by adding
linear independent spin terms with axial
symmetry.
In both cases, we come to the representation of the Hamiltonian (\ref{eq:H}) in the
Bloch-type form.
In particular, the Hamiltonian of the system (1/2,$S$) with spin $S=1$ includes ten
terms and can be written as
\begin{eqnarray}
   \label{eq:Ha}
   {\cal H}&=&B_1s_z+B_2S_z+J(s_xS_x+s_yS_y)+J_zs_zS_z+KS_z^2+K_1(S_x^2+S_y^2)+K_2s_zS_z^2
+D_z(s_xS_y-s_yS_x)
   \nonumber\\
&+&\Gamma[s_x(S_xS_z+S_zS_x)+s_y(S_yS_z+S_zS_y)]
+\Lambda[s_x(S_yS_z+S_zS_y)-s_y(S_xS_z+S_zS_x)],
\end{eqnarray}
where $s_\mu=\sigma_\mu/2$, in which $\sigma_\mu$ are again the Pauli matrices, and
$S_x$, $S_y$ and $S_z$ are the components of spin $S$ (setting $\hbar=1$).

In the Hamiltonian (\ref{eq:Ha}),
$B_1$ and $B_2$ are the $z$-components of an external magnetic field applied to spins
1/2 and 1, respectively;
$J$ and $J_z$ are the exchange Heisenberg constants;
$K$ is the single-ion uniaxial anisotropy;
$K_1$ is the planar single-ion anisotropy;
$K_2$ is the uniaxial two-ion anisotropy;
$D_z$ is the $z$-component of Dzyaloshinskii vector;
$\Gamma$ and $\Lambda$ are respectively symmetric and asymmetric higher-order spin
coupling terms.
It is interesting to note that the antisymmetric DM interaction satisfies axial
symmetry, while the symmetric combination $s_xS_y+s_yS_x$, which
corresponds to the so-called KSEA
interaction \cite{Y20}, does not satisfy.
The number of terms in the Hamiltonian (\ref{eq:Ha}) for systems with $S>1$ increases,
but we will limit ourselves to the same ten terms in order to make comparative study
between cases with different spins $S$.

The Gibbs density matrix is defined as
\begin{equation}
   \label{eq:rhoG}
   \rho=\frac{1}{Z}\exp(-{\cal H}/T),
\end{equation}
where $Z={\rm Tr}\exp(-{\cal H}/T)$ is the partition function and $T$ is the temperature.
Since $\cal H$ has the axial
symmetrical form, this functional relationship guarantees the axial
symmetry form for
the matrix $\rho$.

To find Gibbs density matrix, we first rewrite the Hamiltonian (\ref{eq:Ha}) as an
open matrix.
This allows us to express all the entries of the matrix (\ref{eq:H}) through the
parameters of this Hamiltonian.
Next, it is clear that the energy levels of the system (\ref{eq:H}) are $E_ 0$,
$E_{4S+1}$ and $2S$ additional pairs
\begin{equation}
   \label{eq:Ei}
   E_{k,2S+k}=\frac{1}{2}\big(h_k+h_{2S+k}\pm R_k\big),
\end{equation}
where
\begin{equation}
   \label{eq:Rk}
   R_k=\sqrt{(h_k-h_{2S+k})^2+4|g_k|^2}.
\end{equation}
The partition function $Z=\sum_n\exp(-E_n/T)$ is given by
\begin{equation}
   \label{eq:Z}
   Z=e^{-E_0/T}+e^{-E_{4S+1}/T}+2\sum_{k=1}^{2S}\cosh{\frac{R_k}{2T}}e^{-(h_k+h_{2S+k})/2T}.
\end{equation}
Statistical weights, i.e. density matrix eigenvalues (\ref{eq:p_k}), would be obtained
as
\begin{equation}
   \label{eq:p_i}
   p_0=\frac{1}{Z}e^{-E_0/T},\qquad
	 p_k=\frac{1}{Z}e^{-E_k/T},\qquad p_{2S+k}=\frac{1}{Z}e^{-E_{2S+k}/T},\qquad p_{4S+1}=\frac{1}{Z}e^{-E_{4S+1}/T}.
\end{equation}
Finally,
if the dimension of the matrix $\cal H$ is finite, $\rho$ can be found manually or
obtained on a digital machine by direct symbolic (analytical) calculations.
After this, the results can be extended by induction to systems with arbitrary spin
$S$.
This allows one to get all nonzero matrix elements of the inner block of the density
matrix (\ref{eq:rho}):
\begin{eqnarray}
   \label{eq:akuk}
   a_{k,2S+k}&=&\frac{1}{Z}\Big(\cosh{\frac{R_k}{2T}}\pm\frac{h_{2S+k}-h_k}{R_k}\sinh{\frac{R_k}{2T}}\Big)e^{-(h_k+h_{2S+k})/2T},
   \nonumber\\
	 &&u_k=-\frac{2g_k}{ZR_k}\sinh{\frac{R_k}{2T}}e^{-(h_k+h_{2S+k})/2T}.
\end{eqnarray}
So, the Gibbs density matrix has been found in terms of interaction constants of the
Hamiltonian (\ref{eq:Ha}).

We can now move on to studying the various properties of quantum correlations.


\section{Behavior of quantum correlations}
\label{sec:res}


\subsection{
High-temperature region
}
\label{subsect:high-T}
Let us begin our analysis with the behavior of quantum correlations at high temperatures
$T\to\infty$.
Using formulas (\ref{eq:U0}), (\ref{eq:U1}), (\ref{eq:F0}) and (\ref{eq:F1}), we
extract high-temperature asymptotics of the branches of quantum correlations for the
system (\ref{eq:Ha}).
The results are as follows.

Spin $S=1$:
\begin{equation}
   \label{eq:FU0FU1_T8_S1}
   {\cal F}_0=2\,{\cal U}_0=\frac{J^2+D_z^2+\Gamma^2+\Lambda^2}{3T^2}+O(1/T^3),\quad
\end{equation}

\begin{equation}
   \label{eq:FU0FU1_T8_S1a}
   {\cal F}_1=2\,{\cal U}_1=\frac{3B_1^2+2K_2^2+2(J^2+J_z^2+D_z^2+\Gamma^2+\Lambda^2+2B_1K_2)}{12T^2}+O(1/T^3).
\end{equation}

Spin $S=3/2$:
\begin{equation}
   \label{eq:FU0FU1_T8_S1.5}
   {\cal F}_0=2\,{\cal U}_0=\frac{5(J^2+D_z^2)+12(\Gamma^2+\Lambda^2)}{8T^2}+O(1/T^3),\quad
\end{equation}
\begin{equation}
   \label{eq:FU0FU1_T8_S1.5a}
   {\cal F}_1=2\,{\cal U}_1=\frac{16B_1^2+41K_2^2+20(J^2+J_z^2+D_z^2+2B_1K_2)+48(\Gamma^2+\Lambda^2)}{64T^2}+O(1/T^3).
\end{equation}

Spin $S=2$:
\begin{equation}
   \label{eq:FU0FU1_T8_S2}
   {\cal F}_0=2\,{\cal U}_0=\frac{2(J^2+D_z^2)+3(\Gamma^2+\Lambda^2)}{2T^2}+O(1/T^3),\quad
\end{equation}
\begin{equation}
   \label{eq:FU0FU1_T8_S2a}
   {\cal F}_1=2\,{\cal U}_1=\frac{5B_1^2+34K_2^2+10(J^2+J_z^2+D_z^2+2B_1K_2)+42(\Gamma^2+\Lambda^2)}{20T^2}+O(1/T^3).
\end{equation}

Spin $S=5/2$:
\begin{equation}
   \label{eq:FU0FU1_T8_S2.5}
   {\cal F}_0=2\,{\cal U}_0=\frac{35(J^2+D_z^2)+224(\Gamma^2+\Lambda^2)}{24T^2}+O(1/T^3),\quad
\end{equation}
\begin{equation}
   \label{eq:FU0FU1_T8_S2.5a}
   {\cal F}_1=2\,{\cal U}_1=\frac{48B_1^2+707K_2^2+140(J^2+J_z^2+D_z^2+2B_1K_2)+896(\Gamma^2+\Lambda^2)}{192T^2}+O(1/T^3).
\end{equation}
These expressions show that the high-temperature tails (wings) of quantum correlations
monotonically tend to zero according to the power law $T^{-2}$.
Thus, temperature has a destructive effect on quantum correlations in this region and
completely destroys them in the limit $T\to\infty$.

Next, the $B_2$ field and the single-ion anisotropy strengths $K$ and $K_1$ are absent
here; they manifest themselves in higher terms of high-temperature expansions.
We also see that the main high-$T$ terms ${\cal F}_{0,1}$ are exactly twice as large as
the corresponding terms of ${\cal U}_{0,1}$.

Equations~(\ref{eq:FU0FU1_T8_S1}) - (\ref{eq:FU0FU1_T8_S2.5a}) make it easy to obtain
valuable information, namely 0- or/and 1-branches determine the behavior of quantum
correlations ${\cal U}=\min{\{{\cal U}_0,{\cal U}_1\}}$ and
${\cal F}=\min{\{{\cal F}_0,{\cal F}_1\}}$ in the high-temperature region.

All these features will be important in the following subsections  when analyzing the
behavior of quantum correlations over the entire temperature region.


\subsection{Increasing quantum correlations with temperature}
\label{subsec:XXZBKm}
The parameter space of the studied model (\ref{eq:Ha}) is $R^{10}$.
The space is huge and there is no possibility to present all the obtained graphic
material.
Therefore, we will limit ourselves to only some discovered features of quantum
correlations.

%
\begin{figure}[t]
\begin{center}
\epsfig{file=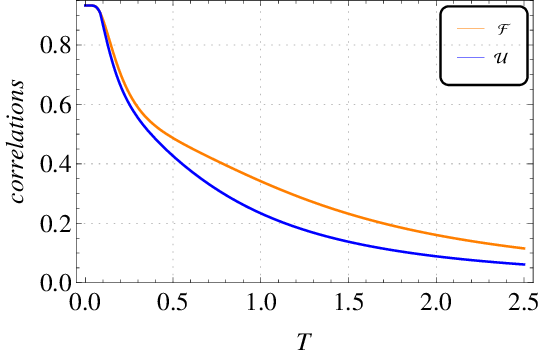,width=5.7cm}
		\put(-135,110){($\rm\bf a$)}
\hspace{1mm}
\epsfig{file=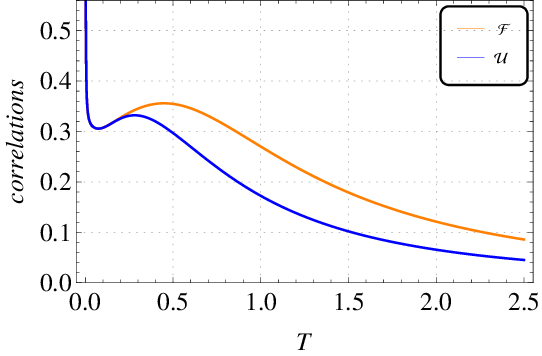,width=5.7cm}
		\put(-135,110){($\rm\bf b$)}
\hspace{1mm}
\epsfig{file=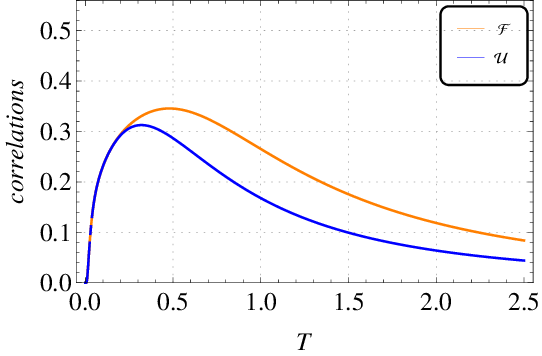,width=5.7cm}
		\put(-135,110){($\rm\bf c$)}
\end{center}
\begin{center}
\caption{
Quantum correlations $\cal F$ and $\cal U$ as functions of temperature $T$ for the
qubit-qutrit model with $B_1=B_2=K=K_1=K_2=D_z=0$, $J=1$, $J_z=-1.7$, $\Lambda=1$, and
$\Gamma=1$~({\bf a}), 0.38 ({\bf b}) and 0.3 ({\bf c}).
}
\label{fig:zzS1ag}
\end{center}
\end{figure}
%

%
\begin{figure}[t]
\begin{center}
\epsfig{file=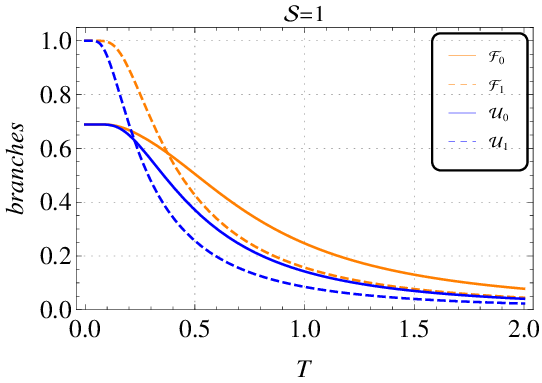,width=6.8cm}
\hspace{3mm}
\epsfig{file=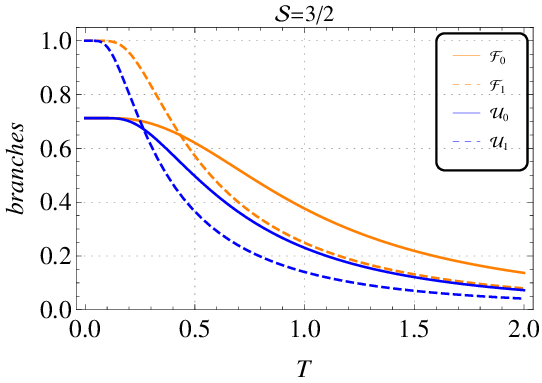,width=6.8cm}
\hspace{3mm}
\epsfig{file=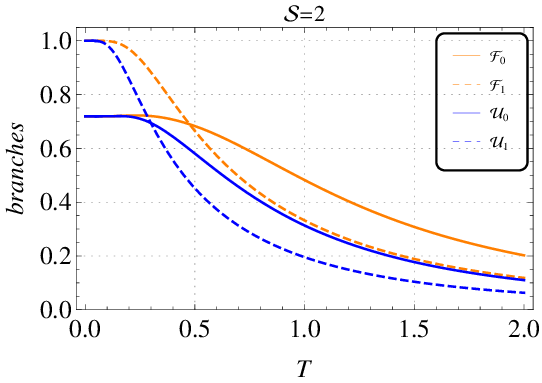,width=6.8cm}
\hspace{3mm}
\epsfig{file=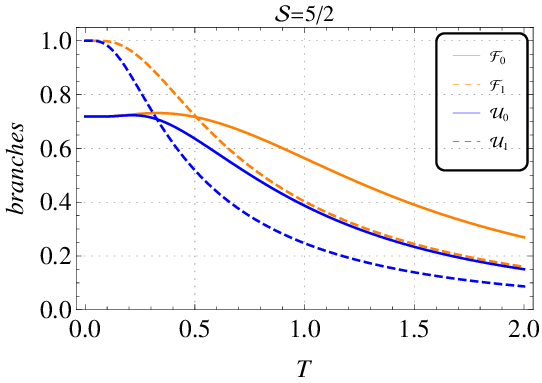,width=6.8cm}
\end{center}
\begin{center}
\caption{
Branches of quantum correlations LQFI and LQU depending on the temperature $T$ for the
model with $B_1=0$, $B_2=-0.8$, $J=1$, $J_z=0.3$
and $K=K_1=K_2=D_z=\Gamma=\Lambda=0$
at the spin lengths $S=1,\,3/2,\,2$ and 5/2.
}
\label{fig:zz17bran}
\end{center}
\end{figure}
%

The behavior at high temperatures has been established above.
More interesting features and peculiarities of quantum correlations are observed in
lower temperature regions.

Let us consider the XXZ model with only $\Gamma$- and $\Lambda$-interactions;
$J=1$, $J_z=-1.7$, $\Lambda=1$, and $\Gamma$ is a variable parameter.
Look at Fig.~\ref{fig:zzS1ag}.
From Fig.~\ref{fig:zzS1ag}a it is clearly seen that at $\Gamma=1$ both correlations LQU
and LQFI monotonically tend to zero with increasing temperature.
Thus, as the system heats up, quantum correlations gradually weaken.
This is in line with our intuitive expectations.

As the interaction $\Gamma$ decreases, monotonicity breaks down, as can be seen from
Fig.~\ref{fig:zzS1ag}b, where $\Gamma=0.38$.
The curves are deformed, minima appear near $T\approx0.076$ and maxima
at $T\approx0.29$ for $\cal U$ and at $T\approx0.47$ for $\cal F$.
It is important to emphasize that both quantum correlations {\em increase} with
an increase in temperature $T$ in the local interval between these minima and maxima.
And this happens despite the fact that, generally speaking, temperature causes
degradation of the coherence and quantum nature of physical systems.

Moreover, temperature can generate nonclassical correlations near the ground state and
then increase them.
This is clearly shown by the curves in Fig.~\ref{fig:zzS1ag}c, which
corresponds to $\Gamma= 0.3$.
In the temperature range from zero to $T\approx0.32$, LQU increases from zero to about
0.31, and LQFI increases from zero to about 0.34 when the temperature varies from zero
to $T\approx0.48$.

%
\begin{figure}[t]
\begin{center}
\epsfig{file=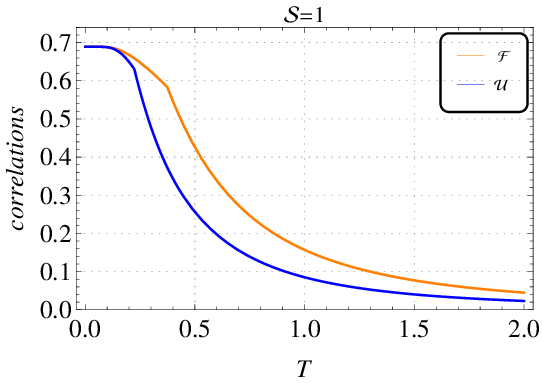,width=6.8cm}
\hspace{3mm}
\epsfig{file=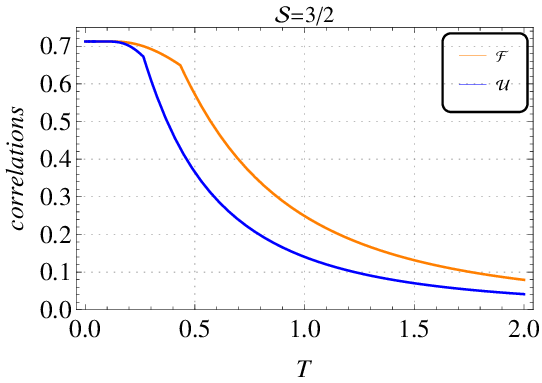,width=6.8cm}
\hspace{3mm}
\epsfig{file=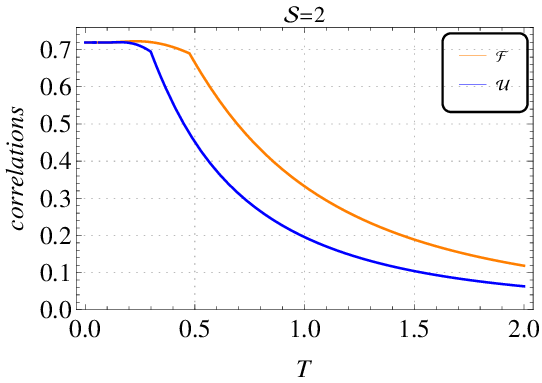,width=6.8cm}
\hspace{3mm}
\epsfig{file=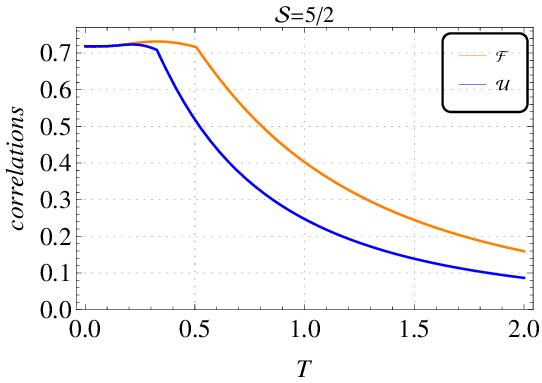,width=6.8cm}
\end{center}
\begin{center}
\caption{
Quantum correlations $\cal F$ and $\cal U$ as functions of temperature $T$ for the
model with $B_1=0$, $B_2=-0.8$, $J=1$, $J_z=0.3$
and $K=K_1=K_2=D_z=\Gamma=\Lambda=0$
at different values of spin length $S$.
}
\label{fig:zz17corr}
\end{center}
\end{figure}
%

Thus, temperature destroys the quantum correlations of a physical system at high
temperatures.
On the other hand, heating the system can lead to a local increase in its quantumness
in the ground state and above it.

Finally, we note that the behavior of LQU and LOFI qualitatively reproduces each other
in all cases presented in Fig.~\ref{fig:zzS1ag}.


\subsection{
Sadden change phenomena
}
\label{subsec:XXZBK}
Quantum correlations LQU and LQFI are determined by Eqs.~(\ref{eq:U0U1}) and
(\ref{eq:F0F1}), respectively.
In general, these functions are piecewise-defined.
The minimization conditions included in their definitions can lead to an
interesting phenomenon, namely, the behavior of quantum correlations can undergo
{\em sharp} changes as the parameters of the system change {\em smoothly}.
In the mathematical literature, such phenomena are called catastrophes \cite{A92}.

Consider the system, in which $B_1=K=K_1=K_2=D_z=\Gamma=\Lambda=0$, i.e. there is
XXZ dimer in an external magnetic field $B_2$.
As follows from Eqs.~(\ref{eq:FU0FU1_T8_S1}) - (\ref{eq:FU0FU1_T8_S2.5a}), 0-branches
are below 1-branches if $|J_z|>1$, and, conversely, above if $|J_z|<1$.

In the absence of a second external field ($B_2=0$), pairs of branches ${\cal U}_0(T)$
and ${\cal U}_1(T)$, as well as ${\cal F}_0(T)$ and ${\cal F}_1(T)$ do not intersect.
For $|J_z|>1$, the quantum correlations ${\cal U}(T)={\cal U}_0(T)$ and
${\cal F}(T)={\cal F}_0(T)$.
If $|J_z|<1$, then the correlations ${\cal U}(T)$ and ${\cal F}(T)$ are completely
determined by the branches ${\cal U}_1(T)$ and ${\cal F}_1(T)$, respectively.
In both cases, the quantum correlations are smooth functions.

The situation begins to change dramatically at $B_2\ne0$.
Figure~\ref{fig:zz17bran} shows the temperature dependencies of branches for different
spin values $S$.
The intersections of branches are clearly observed here.
The ${\cal U}_0(T)$ curves intersect the ${\cal U}_1(T)$ ones at the points with
temperatures 0.215, 0.262, 0.292 and 0.323 for $S=1$, 3/2, 2 and 5/2, respectively.
These temperatures increase with increasing $S$.
Similarly, they are true for the $\cal F$-branches except that the intersects occur at
different temperatures.

On the correlation curves, the above intersection points are visible as fractures
(sharp bends), see Fig.~\ref{fig:zz17corr}.
They are non-analytical points.
The functions ${\cal U}(T)$ and ${\cal F}(T)$ are continuous, but their first
derivatives experience discontinuities of the first kind and therefore these functions
belong to the differentiability class $C^0$.

Experimental detection of singular points and study of the behavior of quantum
correlations near them carry valuable information about the systems under examination.
This information can make it possible to draw conclusions about interaction constants
and understand the role of certain bonds in processes occurring in various substances.

%
\begin{figure}[t]
\begin{center}
\epsfig{file=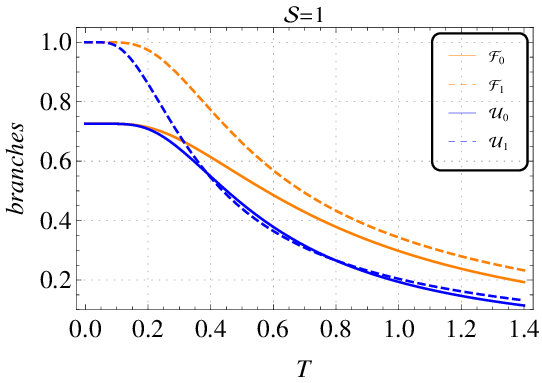,width=6.8cm}
\hspace{3mm}
\epsfig{file=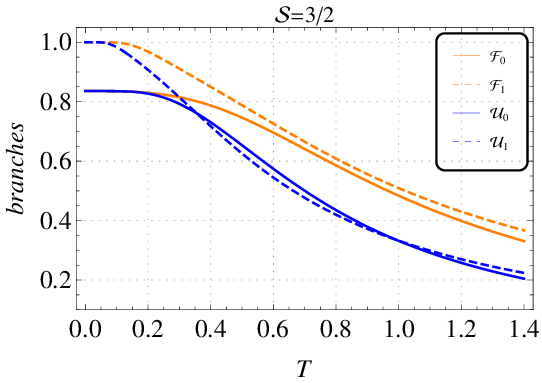,width=6.8cm}
\end{center}
\begin{center}
\caption{
Branches of quantum correlations depending on temperature for the model with
$B_1=-0.65$, $B_2=0.9$, $J=-1.3$, $J_z=-1$, $K=0.4$, $K_2=-1.02$, $D_z=-0.3$ and
$K_1=\Gamma=\Lambda=0$
at spin lengths $S=1$ and 3/2.
}
\label{fig:zS1_a18}
\end{center}
\end{figure}
%

%
\begin{figure}[t]
\begin{center}
\epsfig{file=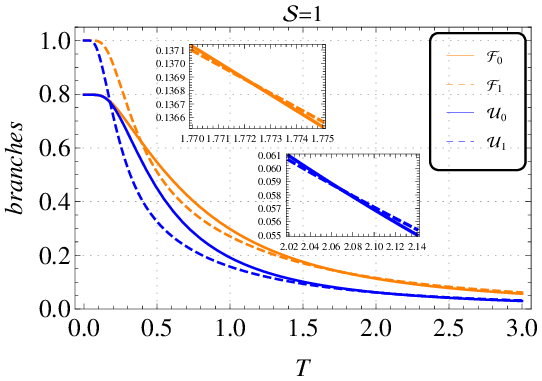,width=7.8cm}
\hspace{3mm}
\epsfig{file=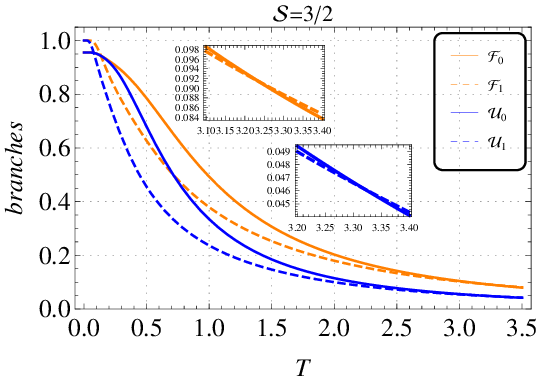,width=7.8cm}
\end{center}
\begin{center}
\caption{
Branches of quantum correlations depending on temperature for the model with
$B_1=-0.65$, $B_2=0.9$, $J=-1.3$, $J_z=-1$, $K=0.4$, $K_2=-0.6$, $D_z=-0.3$ and
$K_1=\Gamma=\Lambda=0$ at spin lengths $S=1$ and 3/2.
Insets show the vicinities of intersection points at higher temperatures.
}
\label{fig:zS1_a15m}
\end{center}
\end{figure}
%

As follows from Figs.~\ref{fig:zz17bran} and \ref{fig:zz17corr}, the above systems
contain one sudden transition.
It is remarkable that a series of such transitions is possible.
Indeed, let
$B_1=-0.65$, $B_2=0.9$, $J=-1.3$, $J_z=-1$, $K=0.4$, $K_2=-1.02$, $D_z=-0.3$ and
$K_1=\Gamma=\Lambda=0$.
As can be seen from Fig.~\ref{fig:zS1_a18}, the $\cal U$-branches (blue curves)
intersect twice and hence the behavior of LQU suddenly changes at two different
temperatures.
It is interesting to note that the second correlation, LQFI, preserves continuity and
smoothness.
This indicates a peculiar hybrid behavior of the non-classical correlation in the given
system.

Finally, let us consider a model with slightly different parameters, namely, $B_1$,
$B_2$, $J$, $J_z$, $K$, $K_1$, $D_z$, $\Gamma$ and $\Lambda$ remain the same,
but $K_2$ is replaced by $-0.6$.
Look at Fig.~\ref{fig:zS1_a15m}.
Here, each pair of 0- and 1-branches intersects at two points.
As a result, when the temperature changes from zero to high values, the quantum
correlations $\cal U$ and $\cal F$ undergo two sudden transitions.
Figure~\ref{fig:zS1_a15m} clearly shows the intersections of the curves at low
temperatures.
The $\cal U$ branches (blue curves) intersect at temperatures 0.164 and 0.06 for $S$=1
and 3/2, respectively.
Similarly for $\cal F$-branches (orange curves): they intersect at temperatures 0.375
and 0.13 for spin lengths $S$=1 and 3/2, respectively.
Unfortunately, at higher temperatures the intersections become less visible, so their
surroundings are shown in the insets.
From these insets it is clear that at high temperatures the behavior of quantum
correlations is determined by 0-branches.
The same conclusion follows from the high-temperature decompositions
(\ref{eq:FU0FU1_T8_S1}) - (\ref{eq:FU0FU1_T8_S1.5a}).

In conclusion of this subsection, we would like to note the following.
Although the LQU and LQFI are discord-type quantum correlations, they are determined
only by two optimal measurement angles, namely zero and $\pi/2$.
Meanwhile, the original (entropic) discord may have an optimal measurement angle
between zero and $\pi/2$.
This leads to sudden transitions, which are described by Landau's theory of phase
transitions of the first and second kind \cite{Y22}.


\section{
Summary and outlook
}
\label{sec:Summ}
In this paper, we have derived analytical formulas (\ref{eq:U0}), (\ref{eq:U1}),
(\ref{eq:F0}) and (\ref{eq:F1}) for the branches of quantum correlations LQU and LQFI.
These formulas turned out to be unexpectedly compact, as a result of which computer
code for them is written very simply.
This actually makes it possible to study the local quantum uncertainty $\cal U$ and the
local quantum Fisher information $\cal F$ for any  spin-$(1/2,S)$ systems in
axisymmetric states.
The state for a given $S$ contains $8S+1$ free parameters.
This allows the investigation of many physically important models.

The derived formulas were then applied to the systems in thermal equilibrium, for which
the behavior of quantum correlations depending on temperature was studied.
The high-temperature asymptotic behavior of quantum correlations was established,
whence it follows that both LQU and LQFI decrease monotonically with increasing $T$
according to the power law $T^{-2}$ and completely disappear at an infinitely high temperature.
On the other hand, quantum correlations can increase in finite (limited) regions with
increasing temperature.
Additionally, we showed that temperature can induce quantum correlations from
an uncorrelated ground state.

Next, we studied the effects of sudden changes in quantum correlations for
different values of $S$.
The performed calculations showed that both LQU and LQFI can experience one or more
such transitions.

In conclusion, we investigated the quantum correlations LQU and LQFI and found that they allow us to track the persistence of quantum correlations at finite temperatures and reveal intricate thermal effects such as correlation enhancement and abrupt transitions. However, it is not possible to state precisely which measure is more efficient for quantifying quantum correlations as it depends on the model and parameters of the system under consideration.

The obtained formulas can be used to study other properties of quantum correlations,
for example, their dynamical behavior.
We believe that axially
symmetric quantum states of qubit-qudit systems will be as useful as X states
for two-qubit models.


\section*{Acknowledgements}
E.I.K. and M.A.Y. were supported in part by a state task, the state
registration number of the Russian Federation is $\#$124013000760-0.
S.H. thanks the School of Particles and Accelerators at the Institute for Research in Fundamental Sciences for their financial support.

\section*{Data availability statement}
All data that support the findings of this study are included within the article.
\section*{Author contributions}
{\bf Saeed~Haddadi}:\ Methodology, Investigation, Writing – review \& editing.
{\bf Elena~I.~Kuznetsova}:\ Calculations, Interpretation of results.
{\bf M.~A.~Yurischev}:\ Writing - original draft, Conceptualization.
Thorough checking of the manuscript was done by all authors.
\section*{Conflict of interest}
The authors declare that no conflicts of interest or personal relationships have
influenced this work.


\section*{ORCID iDs}
Saeed Haddadi \href{https://orcid.org/0000-0002-1596-0763}{https://orcid.org/0000-0002-1596-0763}

Elena I. Kuznetsova \href{https://orcid.org/0000-0002-0053-0023}{https://orcid.org/0000-0002-0053-0023}

M. A. Yurischev \href{https://orcid.org/0000-0003-1719-3884}{https://orcid.org/0000-0003-1719-3884}


\appendix
\section{Diagonalization of the density matrix}
\label{appen:A}
Since the density matrix $\rho$ commutes with the diagonal matrix ${\cal S}_z$, it has
nonzero matrix elements only between states with the same eigenvalue of ${\cal S}_z$.
Here, ${\cal S}_z$ plays a role of ``constant or integral of motion''; in this
connection see, e.g., Chapter~1 in the books~\cite{B64,BJ86}.
As mentioned in Sec.~\ref{sect:rho}, ${\cal S}_z$ has two non-degenerate and $4S$
doubly degenerate eigenvalues.
Therefore, the density matrix (\ref{eq:rho}) actually consists of two
one-by-one and $2S$ two-by-two subblocks.\footnote{
  This will be obvious if to reorder the vectors of the original basis so that the
	$k$th and $(2S+k)$th vectors become neighbors, then the density matrix (\ref{eq:rho})
	takes an explicit quasidiagonal form
$
\hat \rho=[p_0,
      \left(
      \begin{array}{cc}
      a_1&u_1\ \\
      u_1^*&a_{2S+1}\ \\
      \end{array}
      \right),\ldots,
      \left(
      \begin{array}{cc}
      a_{2S}&u_{2S}\ \\
      u_{2S}^*&a_{4S}\ \\
      \end{array}
      \right),
      p_{4S+1}].
$
  This form is implemented for the system $(S,1/2)$.
}
The first and $(4S+1)$th ``subblocks'' have sizes $1\times1$, i.e. they are ready
eigenvalues.
The remaining $2S$ subblocks are $2\times2$-dimensional, and their diagonalization is
easily accomplished by directly calculating the eigenvectors of the corresponding
subblocks (see, for example, \cite{M57}).
Eigenvectors are constructed using the $k$th and $(2S+k)$th pair of vectors of the
original basis.
Indeed, a suitable fragment extracted from the inner $4S\times4S$ block of the density
matrix (\ref{eq:rho}) reads
\begin{equation}
   \label{eq:ak}
	 \left(
      \begin{array}{ccc}
      a_k&\ldots&u_k\ \\
      \vdots&\ddots&\vdots\ \\
      u_k^*&\ldots&a_{2S+k}\ \\
      \end{array}
   \right),\qquad k=1,\ldots,2S
\end{equation}
and then the corresponding eigenvectors are given as
\begin{equation}
   \label{eq:vec_k}
	 |k\rangle=
	 \left(
      \begin{array}{c}
      {\tilde q}_k\ \\
      \vdots\ \\
      {\tilde u}_k^*\ \\
      \end{array}
   \right),\qquad
	 |2S+k\rangle=
	 \left(
      \begin{array}{c}
      {\tilde u}_k\ \\
      \vdots\ \\
      -{\tilde q}_k\ \\
      \end{array}
   \right),\qquad
\end{equation}
where
\begin{equation}
   \label{eq:qk_tilde}
   {\tilde q_k}=q_k/\sqrt{q_k^2+|u_k|^2},\qquad {\tilde u_k}=u_k/\sqrt{q_k^2+|u_k|^2},\qquad
	 q_k=\frac{1}{2}\Big(a_k-a_{2S+k}+\sqrt{(a_k-a_{2S+k})^2+4|u_k|^2}\Big).
\end{equation}
This makes it possible to immediately write down the desired transformation, which
diagonalizes the density matrix (\ref{eq:rho}):
\begin{equation}
   \label{eq:R}
   R=
	 \left(
      \begin{array}{clccccccrc}
      1&\ &\ &\ &\ &\ &\ &\ &\ &\ \\
      \ &{\tilde q}_1&0&\ldots&0&{\tilde u}_1&0&\ldots&0&\ \\
      \ &0&{\tilde q}_2&\ldots&0&0&{\tilde u}_2&\ldots&0&\ \\
      \ &\vdots&\vdots&\ddots&\vdots&\vdots&\vdots&\ddots&\vdots&\ \\
      \ &0&0&\ldots&{\tilde q}_{2S}&0&0&\ldots&{\tilde u}_{2S}&\ \\
      \ &{\tilde u}_1^*&0&\ldots&0&-{\tilde q}_1&0&\ldots&0&\ \\
      \ &0&{\tilde u}_2^*&\ldots&0&0&-{\tilde q}_2&\ldots&0&\ \\
      \ &\vdots&\vdots&\ddots&\vdots&\vdots&\vdots&\ddots&\vdots&\ \\
      \ &0&0&\ldots&{\tilde u}_{2S}^*&0&0&\ldots&-{\tilde q}_{2S}&\ \\
			\ &\ &\ &\ &\ &\ &\ &\ &\ &1
      \end{array}
   \right),
\end{equation}
where ${\tilde q}_k$ and ${\tilde u}_k$ ($k=1,\ldots,2S$) are defined by
Eq.~(\ref{eq:qk_tilde}).
Note that $R$ has the axial
symmetrical structure.
Moreover,
it is clear that $R^\dagger=R$ and $R^\dagger R=I$, so this diagonalizing
transformation is both Hermitian and unitary.

Using Eqs.~(\ref{eq:rho}) and (\ref{eq:R}) we find
\begin{equation}
   \label{eq:rho1}
   \rho^\prime\equiv R\rho R=[p_0,p_1,\ldots,p_{2S}, p_{2S+1},\ldots,p_{4S},p_{4S+1}],
\end{equation}
where $p_1,\ldots,p_{4S}$ are the eigenvalues given by Eq.~(\ref{eq:rho}).
From here and Eq.~(\ref{eq:qk_tilde}), it follows that
\begin{equation}
   \label{eq:q_k}
   q_k=p_k-a_{2S+k}
\end{equation}
and
\begin{equation}
   \label{eq:q_ku2_k}
   {\tilde q}_k^2=\frac{1}{2}\Big(1+\frac{a_k-a_{2S+k}}{p_k-p_{2S+k}}\Big),\qquad
	 {\tilde u}_k^2=\frac{1}{2}\Big(1-\frac{a_k-a_{2S+k}}{p_k-p_{2S+k}}\Big),
\end{equation}
where as before $k=1,\ldots,2S$.


\section{Spin matrices in the diagonal representation of the density matrix}
\label{appen:B}
Let us transform the local Pauli matrices $\sigma_\mu\otimes I_{2S+1}$ ($\mu=x,y,z$) of
the particle $A$ into a diagonal representation of the density operator.
Using Eq.~(\ref{eq:R}), we give these matrices in open form
\begin{equation}
   \label{eq:sxI}
   R(\sigma_x\otimes I_{2S+1})R=
	 \left(
      \begin{array}{cccccccccccc}
      \ &{\tilde u}_1^*&\ &\ &\ &\ &-{\tilde q}_1&\ &\ &\ &\ &\ \\
      {\tilde u}_1&0&{\tilde q}_1{\tilde u}_2^*&\ldots&0&0&0&-{\tilde q}_1{\tilde q}_2&\ldots&0&0&\ \\
      \ &{\tilde q}_1{\tilde u}_2&0&\ldots&0&0&{\tilde u}_1{\tilde u}_2&0&\ldots&0&0&\ \\
      \ &\vdots&\vdots&\ddots&\vdots&\vdots&\vdots&\vdots&\ddots&\vdots&\vdots&\ \\
      \ &0&0&\ldots&0&{\tilde q}_{2S-1}{\tilde u}_{2S}^*&0&0&\ldots&0&-{\tilde q}_{2S-1}{\tilde q}_{2S}&\ \\
      \ &0&0&\ldots&{\tilde q}_{2S-1}{\tilde u}_{2S}&0&0&0&\ldots&{\tilde u}_{2S-1}{\tilde u}_{2S}&0&{\tilde q}_{2S}\\
      -{\tilde q}_1&0&{\tilde u}_1^*{\tilde u}_2^*&\ldots&0&0&0&-{\tilde q}_2{\tilde u}_1^*&\ldots&0&0&\ \\
      \ &-{\tilde q}_1{\tilde q}_2&0&\ldots&0&0&-{\tilde q}_2{\tilde u}_1&0&\ldots&0&0&\ \\
      \ &\vdots&\vdots&\ddots&\vdots&\vdots&\vdots&\vdots&\ddots&\vdots&\vdots&\ \\
      \ &0&0&\ldots&0&{\tilde u}_{2S-1}^*{\tilde u}_{2S}^*&0&0&\ldots&0&-{\tilde q}_{2S}{\tilde u}_{2S-1}^*&\ \\
      \ &0&0&\ldots&-{\tilde q}_{2S-1}{\tilde q}_{2S}&0&0&0&\ldots&-{\tilde q}_{2S}{\tilde u}_{2S-1}&0&{\tilde u}_{2S}^*\\
			\ &\ &\ &\ &\ &{\tilde q}_{2S}&\ &\ &\ &\ &{\tilde u}_{2S}&\
      \end{array}
   \right),
\end{equation}

\begin{equation}
   \label{eq:syI}
   R(\sigma_y\otimes I_{2S+1})R\!=\!\!
	 \left(
      \begin{array}{cccccccccccc}
      \ &-i{\tilde u}_1^*&\ &\ &\ &\ &i{\tilde q}_1&\ &\ &\ &\ &\ \\
      i{\tilde u}_1&0&-i{\tilde q}_1{\tilde u}_2^*&\ldots&0&0&0&i{\tilde q}_1{\tilde q}_2&\ldots&0&0&\ \\
      \ &i{\tilde q}_1{\tilde u}_2&0&\ldots&0&0&i{\tilde u}_1{\tilde u}_2&0&\ldots&0&0&\ \\
      \ &\vdots&\vdots&\ddots&\vdots&\vdots&\vdots&\vdots&\ddots&\vdots&\vdots&\ \\
      \ &0&0&\ldots&0&-i{\tilde q}_{2S-1}{\tilde u}_{2S}^*&0&0&\ldots&0&i{\tilde q}_{2S-1}{\tilde q}_{2S}&\ \\
      \ &0&0&\ldots&i{\tilde q}_{2S-1}{\tilde u}_{2S}&0&0&0&\ldots&i{\tilde u}_{2S-1}{\tilde u}_{2S}&0&-i{\tilde q}_{2S}\\
      -i{\tilde q}_1&0&-i{\tilde u}_1^*{\tilde u}_2^*&\ldots&0&0&0&i{\tilde q}_2{\tilde u}_1^*&\ldots&0&0&\ \\
      \ &-i{\tilde q}_1{\tilde q}_2&0&\ldots&0&0&-i{\tilde q}_2{\tilde u}_1&0&\ldots&0&0&\ \\
      \ &\vdots&\vdots&\ddots&\vdots&\vdots&\vdots&\vdots&\ddots&\vdots&\vdots&\ \\
      \ &0&0&\ldots&0&-i{\tilde u}_{2S-1}^*{\tilde u}_{2S}^*&0&0&\ldots&0&i{\tilde q}_{2S}{\tilde u}_{2S-1}^*&\ \\
      \ &0&0&\ldots&-i{\tilde q}_{2S-1}{\tilde q}_{2S}&0&0&0&\ldots&-i{\tilde q}_{2S}{\tilde u}_{2S-1}&0&-i{\tilde u}_{2S}^*\\
			\ &\ &\ &\ &\ &i{\tilde q}_{2S}&\ &\ &\ &\ &i{\tilde u}_{2S}&\
      \end{array}
   \right)
\end{equation}

and

\begin{equation}
   \label{eq:szI}
   R(\sigma_z\otimes I_{2S+1})R=
	 \left(
      \begin{array}{clccccccrc}
      1&\ &\ &\ &\ &\ &\ &\ &\ &\ \\
      \ &{\tilde q}_1^2-|{\tilde u}_1|^2&0&\ldots&0&2{\tilde q}_1{\tilde u}_1&0&\ldots&0&\ \\
      \ &0&{\tilde q}_2^2-|{\tilde u}_2|^2&\ldots&0&0&2{\tilde q}_2{\tilde u}_2&\ldots&0&\ \\
      \ &\vdots&\vdots&\ddots&\vdots&\vdots&\vdots&\ddots&\vdots&\ \\
      \ &0&0&\ldots&{\tilde q}_{2S}^2-|{\tilde u}_{2S}|^2&0&0&\ldots&2{\tilde q}_{2S}{\tilde u}_{2S}&\ \\
      \ &2{\tilde q}_1{\tilde u}_1^*&0&\ldots&0&-{\tilde q}_1^2+|{\tilde u}_1|^2&0&\ldots&0&\ \\
      \ &0&2{\tilde q}_2{\tilde u}_2^*&\ldots&0&0&-{\tilde q}_2^2+|{\tilde u}_2|^2&\ldots&0&\ \\
      \ &\vdots&\vdots&\ddots&\vdots&\vdots&\vdots&\ddots&\vdots&\ \\
      \ &0&0&\ldots&2{\tilde q}_{2S}{\tilde u}_{2S}^*&0&0&\ldots&-{\tilde q}_{2S}^2+|{\tilde u}_{2S}|^2&\ \\
			\ &\ &\ &\ &\ &\ &\ &\ &\ &-1
      \end{array}
   \right).
\end{equation}
Unlike the first two matrices, the latter has the axial
symmetrical form.




\end{document}